\documentclass[aps,prb,groupedaddress,showpacs,twocolumn,superscriptaddress]{revtex4}
\usepackage{graphicx}
\usepackage{amsmath}
\usepackage{amssymb}
\usepackage{bbm}
\usepackage{xspace}
\usepackage{color}
\usepackage{footnote}
\usepackage{hyperref}
\usepackage{subfigure}

\newcommand{\fig}[1]{Fig.\thinspace{}\ref{#1}}
\newcommand{\eq}[1]{Eq.\thinspace{}(\ref{#1})}
\newcommand{\eqs}[1]{Eqs.\thinspace{}(\ref{#1})}
\newcommand{\se}{Sec.\@\xspace}
\newcommand{\tcite}[1]{Ref.~\onlinecite{#1}}
\newcommand{\tcites}[1]{Refs.~\onlinecite{#1}}

\newcommand{\iim}{\Im{}\,}
\newcommand{\atot}{A_{\uparrow\downarrow}}
\newcommand{\nag}{{\phantom{\dagger}}}

\def\ket#1{\mathinner{|{#1}\rangle}}
\def\braket#1{\mathinner{\langle{#1}\rangle}}
\def\Braket#1{\mathinner{\left<{#1}\right>}}

\newcommand{\pw}{0.85}

\begin{document}

\title{Nonequilibrium Kondo effect in a magnetic field: Auxiliary master equation approach}

\author{Delia M. Fugger}
\email{delia.fugger@tugraz.at}
\affiliation{Institute of Theoretical and Computational Physics, Graz University of Technology, Petersgasse 16/II, A-8010 Graz, Austria}
\author{Antonius Dorda}
\affiliation{Institute of Theoretical and Computational Physics, Graz University of Technology, Petersgasse 16/II, A-8010 Graz, Austria}
\author{Frauke Schwarz}
\affiliation{Physics Department, Arnold Sommerfeld Center for Theoretical Physics, and Center for NanoScience,
Ludwig-Maximilians-Universit\"{a}t, Theresienstra{\ss}e 37, D-80333 Munich, Germany}
\author{Jan von Delft}
\affiliation{Physics Department, Arnold Sommerfeld Center for Theoretical Physics, and Center for NanoScience,
Ludwig-Maximilians-Universit\"{a}t, Theresienstra{\ss}e 37, D-80333 Munich, Germany}
\author{Enrico Arrigoni}
\affiliation{Institute of Theoretical and Computational Physics, Graz University of Technology, Petersgasse 16/II, A-8010 Graz, Austria}

\begin{abstract}
We study the single-impurity Anderson  model out of equilibrium under the influence of a bias voltage $\phi$ and a magnetic field  $B$. 
We investigate the interplay between the shift ($\omega_B$) of the Kondo peak in the spin-resolved density of states (DOS) and the one ($\phi_B$) of the conductance anomaly.
In agreement with experiments and previous theoretical calculations we find that, while the latter displays a rather linear behavior with an almost constant slope as a function of $B$ down to the Kondo scale, the DOS shift first features a slower increase reaching the same behavior as $\phi_B$  only for $|g| \mu_B B \gg k_B T_K$. 

Our auxiliary master equation approach yields highly accurate nonequilibrium results for the DOS and for the conductance all the way from within the Kondo up to the charge fluctuation regime, showing excellent agreement with a recently introduced  scheme based on a combination of numerical renormalization group with time-dependent density matrix renormalization group.
\end{abstract}

\maketitle

\section{Introduction}
\label{sec:intro}

Since its discovery almost one century ago, the Kondo effect has been measured in many physical systems ranging from bulk materials to nanostructures.
The latter are especially attractive to study, because the parameters controlling the effect can be precisely tuned in the laboratory.
There is a variety of experiments on nanowires,\cite{ra.bu.94,kr.sh.11,kr.sh.12} two-dimensional electron gases confined in heterostructures, \cite{go.go.98,go.sh.98} carbon nanotubes~\cite{fe.ar.17} and also organic molecules, \cite{zh.ka.13} to mention a few.
Whereas a finite temperature and a bias voltage to probe the effect are perturbations that naturally arise in these experiments and should therefore be studied, it is also interesting to study the effect of an additional magnetic field. 

It is known from these experiments that upon introducing a Zeeman magnetic field $B$ the zero-bias conductance anomaly (i.e. the peak of the conductance $G$ as a function of bias voltage $\phi$) splits into two peaks located at $\pm \phi_B$, where $\phi_B$ increases almost linearly with $B$. \cite{ra.bu.94,am.ge.05,kr.sh.11,kr.sh.12} 
Theoretical calculations~\cite{me.wi.93,ba.us.10,he.ba.05,ro.pa.03,kr.sh.11,re.pl.14,ande.08,sm.gr.13} confirm this behavior showing an essentially constant slope, $e \phi_B \approx |g| \mu_B B$, almost all the way down to the point where the splitting disappears at $|g| \mu_B B \sim k_B T_K$, where $T_K$ is the Kondo temperature that characterizes the width of the zero-bias anomaly at zero temperature and zero field.
At the same time, the magnetic field produces a similar split in the total impurity density of states (spectral function), which again starts developing for magnetic fields of the order of the Kondo scale, and which corresponds to a shift $\pm \hbar \omega_B$ in the spin-resolved impurity density of states.
However, in contrast to $\phi_B$, this shift does not show the same strictly linear behavior.
Accurate calculations based on Bethe ansatz and the numerical renormalization group (NRG)~\cite{mo.we.00,cost.00,cost.03} show that $\omega_B$ is initially smaller, starting as $\hbar \omega_B \approx \frac{2}{3} |g| \mu_B B$ and reaching $|g| \mu_B B$  for $|g| \mu_B B \gg k_B T_K$ (up to logarithmic corrections~\cite{ro.co.03}). 
Notice that less sophisticated equations of motion approaches~\cite{me.wi.92} yield instead a constant slope  of $\omega_B$ as well.
On the other hand, the different behavior of $\omega_B$ and $\phi_B$ is in contradiction with the simple expectation~\cite{me.wi.92} that the enhancement of the conductance should occur when the  chemical potential difference reaches the splitting in the spectral function.
Kondo physics out of equilibrium is a challenging issue from the theoretical point of view and it is hard to obtain accurate results for both the spectral function and the conductance for voltages beyond the linear-response regime, most nonequilibrium steady-state approaches being perturbative or their accuracy being uncontrolled.

In this paper, we investigate the single-impurity Anderson  model (SIAM) in the presence of both a magnetic field $B$ and a finite bias voltage $\phi$. 
We adopt the recently introduced auxiliary master equation approach (AMEA), which has been shown to produce very accurate results for spectral functions and current characteristrics both in as well as out of equilibrium.\cite{do.ga.15} 
To confirm the accuracy of our results we compare them with the ones obtained within a hybrid method that combines NRG with the time-dependent density matrix renormalization group (tDMRG)~\cite{sc.we.17u} to address quantum impurities out of equilibrium. 
The two approaches compare excellently (see \fig{fig:comp}) also at zero bias voltage, where we directly compare the spectral function with NRG.
Our results confirm the different behavior of $\omega_B$ and $\phi_B$, showing that there is no incompatibility.
We also evaluate the magnetization in the high and low field limit, confirming the presence of a plateau at high fields for bias voltages $e \phi \lesssim |g| \mu_B B$ observed in previous theoretical results.\cite{ro.pa.03}

This work is organized as follows:
In \se \ref{sec:modelmethod} the model and the solution method are described.
We start with an introduction to the model, \se \ref{sec:model}, followed by a part about Keldysh Green's functions, \se \ref{sec:Keldysh}.
Then the general idea of AMEA and the solution method are sketched, \se \ref{sec:method}.
In \se \ref{sec:frauke} the hybrid NRG-tDMRG method, which we use for comparison, is described. 
\se \ref{sec:results} contains the results and \se \ref{sec:conclusion} a summary and our conclusions.

\section{Model and Method}
\label{sec:modelmethod}
\subsection{Model}
\label{sec:model}
We study the single-impurity Anderson model (SIAM) in a magnetic field and out of equilibrium. 
Throughout this paper we use units of $\hbar=e=k_B=\mu_B |g|=1$ and $\Gamma=1$, see \eqs{eq:hybstrength_flatband} and \eqref{eq:hybstrength}.
The model is described by the following hamiltonian,
\begin{equation}
  H =  H_{\mathrm{imp}} +  H_{\mathrm{leads}} +  H_{\mathrm{coup}}\,.
 \label{eq:H_parts}
\end{equation}
$H_{\mathrm{imp}}$ is the hamiltonian of the impurity. 
It is a single-site Hubbard hamiltonian with a spin-dependent on-site energy, accounting for the magnetic field,
\begin{equation}
 H_{\mathrm{imp}} = \sum_{\sigma\in\{\uparrow,\downarrow\}} \varepsilon_{f \sigma}^\nag f^\dagger_{\sigma} f_{\sigma}^\nag + U n_{f\uparrow}n_{f\downarrow}\,,
 \label{eq:H_imp}
\end{equation}
with $\varepsilon_{f \sigma} = -\tfrac{1}{2} \left( U + \sigma B \right)$. $f^{(\dagger)}_{\sigma}$ is the fermionic annihilation (creation) operator at the impurity for spin $\sigma$, $n_{f\sigma} = f^\dagger_{\sigma} f_{\sigma}^\nag$, $U$ is the interaction strength and $B$ the magentic field.
The on-site energy $\varepsilon_{f \sigma}$ is chosen such that the system is particle-hole symmetric at $B=0$.
The impurity is connected to two leads described by
\begin{equation}
 H_{\mathrm{leads}} = \sum_{\lambda \in \{L,R\}} \sum_{k \sigma} \varepsilon_{\lambda k}^\nag d^\dagger_{\lambda k \sigma}d^\nag_{\lambda k \sigma} \,.
 \label{eq:H_res}
\end{equation}
$d^{(\dagger)}_{\lambda k \sigma}$ is the annihilation (creation) operator for electrons with spin $\sigma$ in lead $\lambda \in \{L,R\}$ at level $k$ (out of $N$ energy levels); $\varepsilon_{\lambda k}$ is the energy of level $k$. 
The leads have different chemical potentials $\mu_\lambda$, realizing a bias voltage $\phi = \mu_R-\mu_L$ across the impurity.
The hamiltonian mediating the coupling between the impurity and the leads is given by
\begin{equation}
 H_{\mathrm{coup}} = \frac{1}{\sqrt{N}} \sum_{\lambda \in \{L,R\}} t'_\lambda  \sum_{ k \sigma} \left( d^\dagger_{\lambda k \sigma}f^\nag_{\sigma} + \mathrm{H.c.} \right) \,
 \label{eq:H_coup}
\end{equation}
with a symmetric hopping $t'_L = t'_R$. 
We assume that $H_{\mathrm{leads}}$ produces a flat density of states (DOS) $\rho_\lambda(\omega)$ in the disconnected leads with a bandwidth of $2D$,
\begin{equation}
 \rho_\lambda(\omega) = \frac{1}{2D} \Theta(D-|\omega|) \,,
 \label{eq:flat_DOS}
\end{equation}
where $\Theta$ is the Heaviside step function. 
In this flat-band model the hybridization strength $\Gamma$, defined in \eq{eq:hybstrength}, is given by,
\begin{equation}
 \Gamma = \frac{\pi}{2D}\left( {t'_L}^2 + {t'_R}^2 \right)\,.
 \label{eq:hybstrength_flatband}
\end{equation}
Using $\Gamma=1$ as unit of energy yields $t'_\lambda = \sqrt{\frac{D}{\pi}}$ for the hopping to the leads.
Throughout this paper we take $D=10$.

We furthermore use the following definition of the Kondo temperature $T_K$,
\begin{equation}
 G(T=T_K,\phi=0) = \frac{1}{2}G_0\,,
 \label{eq:kondotemp}
\end{equation}
at $B=0$. $G$ is the linear-response differential conductance, \eq{eq:dcurrlin}, $G_0 = G(T=0,\phi=0) = 1/\pi$.

\subsection{Keldysh Green's functions}
\label{sec:Keldysh}
While there is only one independent Green's function in equilibrium, there are two in nonequilibrium: 
The retarded and the Keldysh Green's function, $G^R$ and $G^K$, e.g., are independent of each other.
At finite magnetic field they are furthermore different for both spin kinds.
In steady state, when the system is time-translation invariant, they are defined as
\begin{equation}
\begin{split}
  G^R_\sigma(t) &= -i \Theta(t) \Braket{ \left\{ f_{\sigma}^\nag(t),  f_\sigma^\dagger \right\}  } \,, \\
  G^K_\sigma(t) &= -i  \Braket{ \left[ f_{\sigma}^\nag(t),  f_{\sigma}^\dagger \right]  } \,,
\end{split}
\label{eq:GFs_t}
\end{equation}
and in Fourier space,
\begin{equation}
 G^\alpha_\sigma(\omega) = \int G^\alpha_\sigma(t) \exp(i\omega t) \,dt \,,
\end{equation}
with $\alpha \in \{R,K\}$. Upon introducing the Keldysh contour, these Green's functions can be arranged in a matrix structure, according to
\begin{equation}
   \underline{G}_\sigma(\omega) = \begin{pmatrix} G^R_\sigma(\omega) & G^K_\sigma(\omega) \\ 0 & G^A_\sigma(\omega) \end{pmatrix}\,,
  \label{eq:negf}
\end{equation}
where the advanced Green's function is related to the retarded one by $G^A_\sigma(\omega) = G^R_\sigma(\omega)^\dagger$.
In this way, the familiar form of Dyson's equation is maintained, 
\begin{equation}
\begin{split}
\underline{G}^{-1}_\sigma(\omega) &=\underline{g}^{-1}_{0 \sigma}(\omega) - \underline{\Delta}(\omega)-\underline{\Sigma}(\omega) \\
		  &=\underline{G}^{-1}_{0 \sigma}(\omega) -\underline{\Sigma}(\omega)\,.
\end{split}
\label{eq:dyson}
\end{equation}
$\underline{G}_\sigma(\omega)$ is the full interacting Green's function of the impurity connected to the leads, $\underline{g}_{0 \sigma}(\omega)$ is the noninteracting Green's function of the disconnected impurity, $\underline{\Delta}(\omega)$ is the hybridization of the impurity by the leads and $\underline{\Sigma}(\omega)$ accounts for the interaction at the impurity.
The noninteracting Green's functions are combined to $\underline{G}_{0 \sigma}(\omega)=\underline{g}^{-1}_{0 \sigma}(\omega) - \underline{\Delta}(\omega)$.
The hybridization function is given by
\begin{equation}
 \underline{\Delta}(\omega) = \sum_\lambda {t'_\lambda}^2 \underline{g}_\lambda(\omega) \,,
 \label{eq:Dhyb}
\end{equation} 
where $\underline{g}_\lambda(\omega)$ is the (noninteracting) Green's function of the decoupled leads.
Since these are in equilibrium, its components obey the fluctuation-dissipation theorem, 
\begin{equation}
g^K_\lambda(\omega) = 2\pi i\left(2 f_\lambda(\omega,T) -1 \right) \rho_\lambda(\omega)\,,
 \label{eq:FDT}
\end{equation}
where $f_\lambda = \left[\,\exp{[(\omega-\mu_\lambda)}/T]+1 \right]^{-1}$ denotes the Fermi function at temperature T and chemical potential $\mu_\lambda$.
The DOS in the leads is connected to $g^R_\lambda(\omega)$, 
\begin{align}
 \rho_\lambda(\omega) = -\frac{1}{\pi} \iim g^R_\lambda(\omega).
 \label{eq:Aw}
\end{align}
Therefore in equilibrium only one independent Green's function persists.
The hybridization strength $\Gamma$ is defined, using \eq{eq:Dhyb},
\begin{equation}
 \Gamma = -\iim \Delta^R(\omega=0).
 \label{eq:hybstrength}
\end{equation}

Given the full interacting Green's function at the impurity, the spin-resolved and total spectral functions are calculated as
\begin{equation}
  A_\sigma(\omega) = -\frac{1}{\pi} \iim G^R_\sigma(\omega),~~~~~
  \atot = \frac{1}{2} \left( A_\uparrow + A_\downarrow \right).
  \label{eq:specs}
\end{equation}
The current across the impurity is determined via the Meir-Wingreen formula.\cite{me.wi.92}
In case of a bias-independent lead DOS with $\rho_L(\omega) = \rho_R(\omega)$, such as \eqref{eq:flat_DOS}, it reduces to~\cite{ha.ja}
%
\begin{equation}
\label{eq:curr}
 j = \int \atot(\omega) \, \gamma(\omega) \left( f_R(\omega,T) - f_L(\omega,T)\right) d\omega \,,
\end{equation} 
where $\gamma(\omega) = -\Im \Delta^{R}(\omega)$. 
In linear-response the differential conductance $G=\frac{\partial j}{\partial \phi}$ is calculated from \eqref{eq:curr} as
\begin{equation}
\label{eq:dcurrlin}
 G = \int \atot(\omega) \, \gamma(\omega) \left( - \frac{\partial}{\partial \omega} f(\omega,T) \right) \, d\omega \,,
\end{equation} 
where $f = f_L|_{\mu_L=0} = f_R|_{\mu_R=0}$ is the Fermi function at zero bias.
In the general case, we calculate the differential conductance from finite current differences using three-point Lagrange polynomials to approximate the derivative.

\subsection{Method}
\label{sec:method}
We here present a short sketch of the auxiliary master equation approach (AMEA) used in this paper. 
For more details, we refer to \tcites{do.nu.14,do.ga.15,do.so.17,ar.kn.13}. 
The idea is to map the physical system described by \eqref{eq:H_parts} to a finite and open auxiliary system that has almost the same hybridization at the impurity as the original one \eqref{eq:Dhyb} and thereby maintains the impurity physics, which we are interested in.
The auxiliary system consists of a small number of $N_B$ bath sites connected to Markovian environments and its dynamics is governed by a Lindblad master equation.
The parameters in this equation are determined to achieve a corresponding auxiliary hybridization function $\underline{\Delta}_{aux}(\omega)$ such that $\underline{\Delta}_{aux}(\omega) \approx \underline{\Delta}(\omega)$ as accurately as possible, cf. \onlinecite{do.so.17}. 
The physical hybridization function $\underline{\Delta}$ is calculated from the given lead DOS, \eq{eq:flat_DOS}, using \eqs{eq:Dhyb}-\eqref{eq:Aw} and the Kramers-Kronig relation that links the real and imaginary part of a Green's function.
The auxiliary hybridization function $\underline{\Delta}_{aux}$ can be calculated for a general set of bath parameters by solving a noninteracting Lindblad problem, see, e.g. \tcites{pros.08,pros.10,do.nu.14, do.so.17}.
The determination of these parameters and thus the mapping to the physical system is carried out with a parallel tempering algorithm. \cite{do.so.17}
The resulting Lindblad equation is solved by using matrix product states (MPS) and the time evolving block decimation algorithm (TEBD), as described in \onlinecite{do.ga.15}.
Since the auxiliary Lindblad system is essentially exactly solvable, the approximation of the method lies in the difference between $\underline{\Delta}_{aux}(\omega)$ and $\underline{\Delta}(\omega)$. 
As shown in \tcite{do.so.17}, this difference vanishes exponentially upon increasing $N_B$. 
Therefore, a moderate number of bath sites ($N_B \approx 14-20$) is sufficient to reach the  accuracy required in the present paper.

The results we present here are in the steady state, which is determined via time evolution and formally reached with $t \to \infty$.
The Green's functions are also calculated in the time domain, starting from the steady state; they are continued to large times by linear prediction and then subjected to a Fourier transformation.

The bias voltage is realized by shifting the chemical potentials in the leads symmetrically with respect to each other, $\mu_R = -\mu_L = \frac{\phi}{2}$. 
Note that for each bias voltage a new $\underline{\Delta}_{aux}$ has to be determined, since $\phi$ enters the Keldysh part of the hybridization function.

The calculations for $B<1$, $\phi < 1.8$ and $B>1$, $\phi < 2.1$ are with $N_B = 20$ bath sites; for all other parameters $N_B = 14$ is sufficient.
For the subsequent TEBD calculation we restrict the fit to nearest neighbour couplings.

All results shown in this paper are for the symmetric SIAM, $t'_L = t'_R$. 
Note that the extension to the non-symmetric model is simple and straightforward.

\subsection{Comparison to NRG-tDMRG quench calculations}
\label{sec:frauke}
We compare our data to results obtained in a hybrid NRG-tDMRG quench setup which is described in \tcite{sc.we.17u}. 
While AMEA treats the impurity model as a truly open quantum system in the sense of a Lindblad master equation, for ``small enough'' time scales $t$ one can equally well consider quenches in a closed quantum system. \cite{bo.sa.08,he.fe.09} 
Starting with an initial state in which the two leads are in thermal equilibrium, but held at different chemical potential, standard Hamiltonian time evolution will drive the system towards its ``steady state'' until at some point in time finite-size effects set in. 
For the SIAM one faces the difficulty that the different energy and time scales inherent in the model have to be handled with care. 
The hybrid NRG-tDMRG approach presented in \tcite{sc.we.17u} meets this challenge by exploiting the fact that energy scales outside the transport window, where $f_L(\omega,T)\approx f_R(\omega,T)$, are effectively in equilibrium. 
Thus, they can be traced out using the numerical renormalization group (NRG).\cite{wi.ke.75} 
Subsequently, the non-equilibrium processes arising on the energy scale of the  transport window are treated within this renormalized setup using a tDMRG~\cite{vida.04,da.ko.04,wh.fe.04,scho.11} quench. 
Both methods, NRG and tDMRG, are implemented based on MPS. 

For the high-energy range outside the transport window a logarithmic discretization is used, while the transport window itself is discretized linearly.
After mapping the problem onto a chain, the Hamiltonian of the first part of this chain, which represents the high-energy modes, can be diagonalized using NRG.
This  yields a truncated effective low-energy basis for this part of the system, which can be seen as the local state space of a renormalized impurity (RI).
This RI is coupled to the remainder of the leads, which corresponds to the energy range of the transport window and therefore has an effective bandwidth set by voltage and temperature.
The quench is initialized with a  state $\ket{\Psi}=\ket{\psi_{\text{ini}}}\otimes\ket{\Omega}$, where $\ket{\psi_{\text{ini}}}$ lies in the ground state sector of the RI and $\ket{\Omega}$ is the thermal state of the remaining part of the leads at different chemical potential and decoupled from the RI. This state is time-evolved using tDMRG. The relevant time scale for this quench is given by the size of the transport window.

To further simplify the MPS calculation, the leads are described in the form suggested by the thermofield approach, \cite{ta.um.75,das.00,ve.in.15} in which the thermal state $\ket{\Omega}$ is a pure quantum state, and, even more advantageously, a simple product state on the MPS chain. 
This implies that the time evolution of the tDMRG quench is started with a product state and, hence, with lowest possible entanglement. 

In practice, the time evolution is typically limited, due to the entanglement growth, before finite size effects set in. 
So far, the approach has only been used to calculate expectation values, because the determination of spectral functions would need far more numerical resources. 
For all data points with $\phi>0.14D$ there was no need to use NRG, because the transport window is of similar size as the full bandwidth. 
For high voltages convergence was achieved only in the current and not in the magnetization. 
However, the time dependence of the dot's occupation, $\braket{n_{\sigma}}(t)$, follows an exponential decay such that one can extrapolate to the steady-state value.

\section{Results}
\label{sec:results}

Our approach allows for an accurate solution of the model in and out of equilibrium, below, but also above the energy scale $T_K$, so as to take into account the influence of charge excitations and of the Hubbard bands.  
At the same time, below $T_K$ and in equilibrium our results show a remarkable agreement of the spectral function with NRG up to intermediate values of $U/\Gamma \lesssim 6$, see \fig{fig:spec_eq_comp} and \tcite{do.ga.15}. 
Here we want to study the behavior and interplay of the spectral function and the differential conductance in the presence of a finite Zeeman magnetic field $B$ and bias voltage $\phi$. 
In particular, we focus on the shift of the Kondo and of the zero-bias peak. 

\begin{center}
\begin{figure}[htp]
\subfigure[]{
  \label{fig:spec_eq_spin}
  \begin{minipage}[b]{\pw\columnwidth}
    \centering \includegraphics[width=1\textwidth]{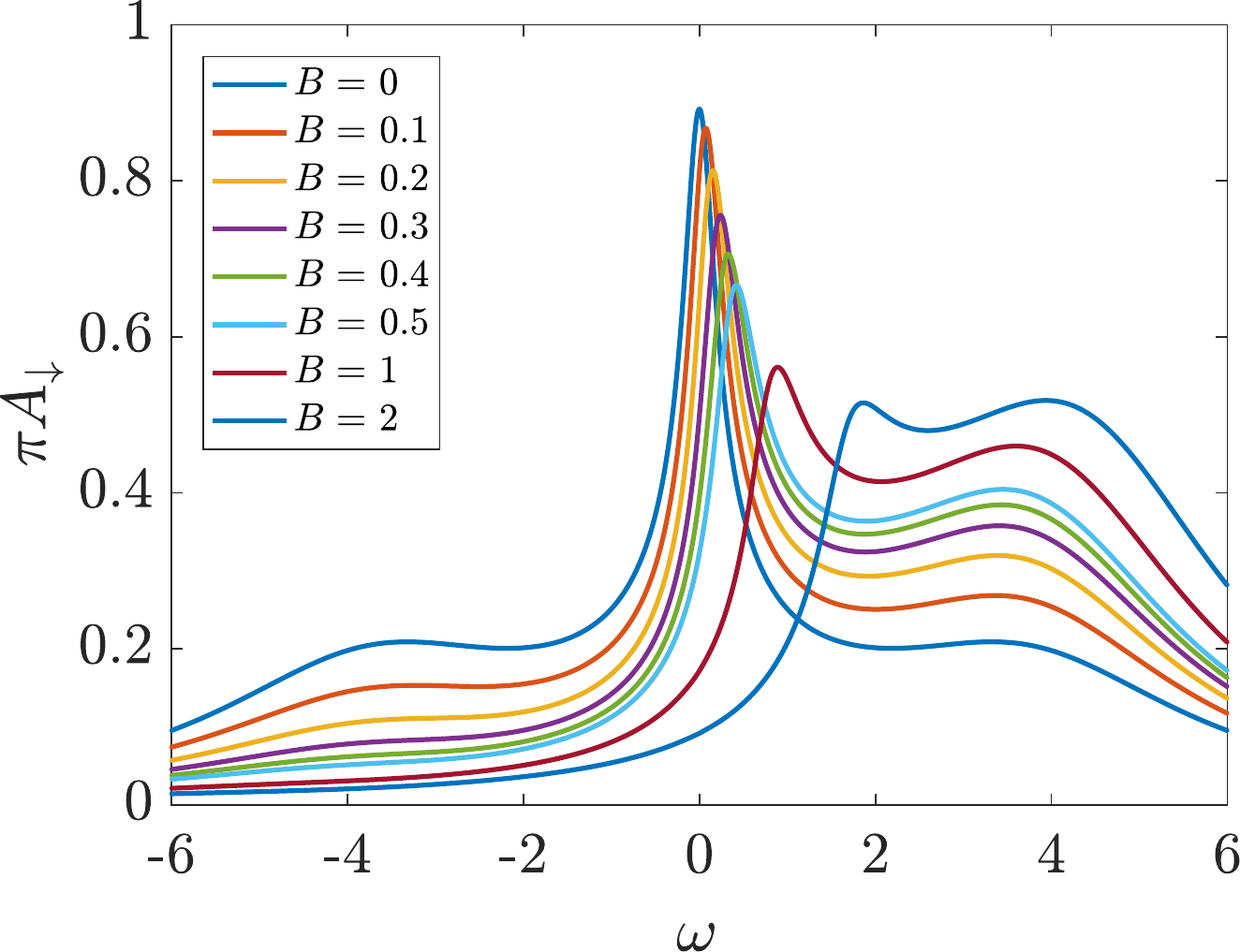}
  \end{minipage}
}\\
\subfigure[]{
  \label{fig:spec_eq_symm}
  \begin{minipage}[b]{\pw\columnwidth}
    \centering \includegraphics[width=1\textwidth]{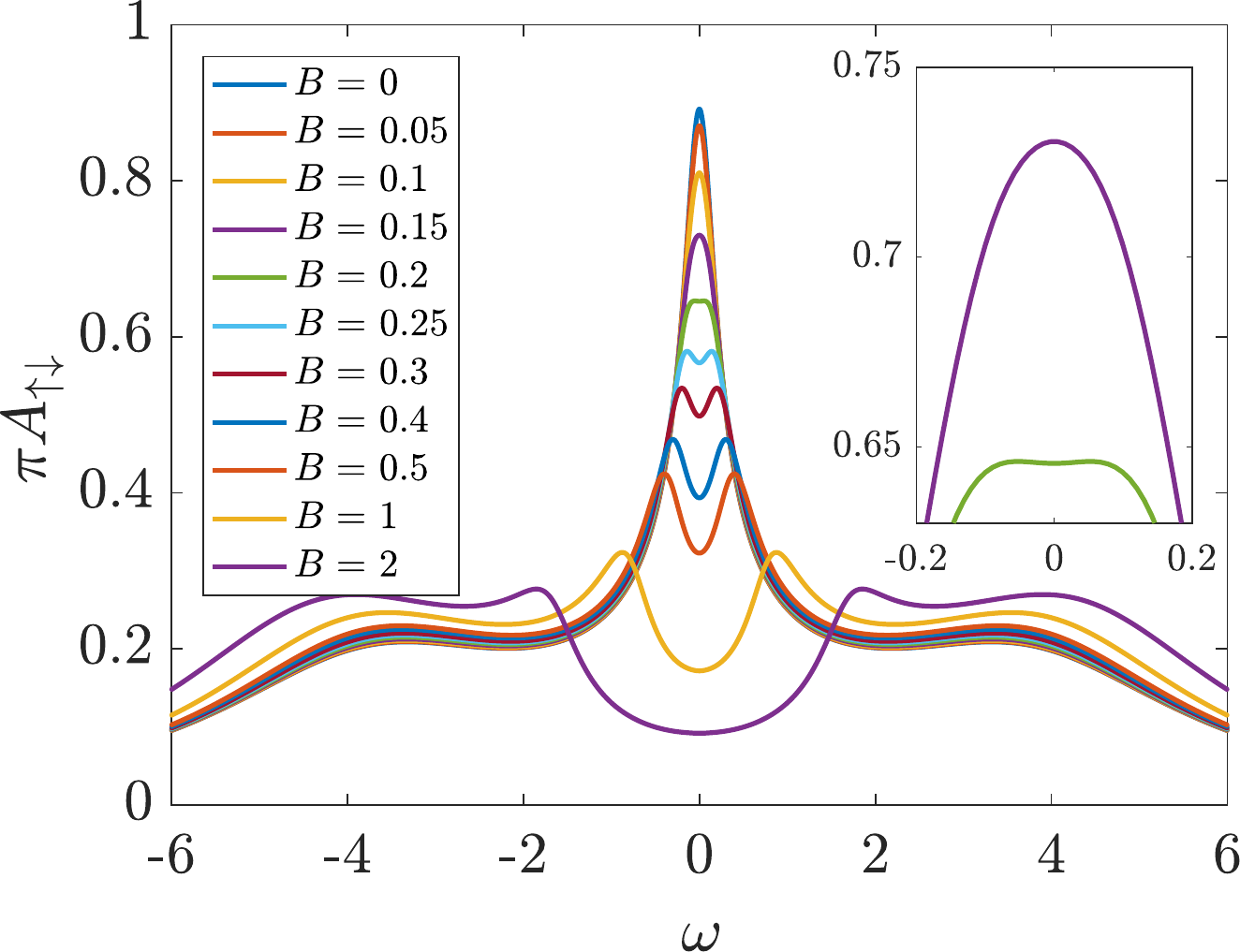}
  \end{minipage}
}\\
\caption{
Equilibrium ($\phi=0$) spin-resolved $A_\downarrow(\omega)$ \subref{fig:spec_eq_spin} and total $\atot(\omega)$ \subref{fig:spec_eq_symm} impurity spectral function for different magnetic fields $B$ and for $U=6\Gamma$ and $T=0.05 \Gamma/k_B \approx T_K/4$.
Note that $B$ is in units of $\Gamma / \left(|g| \mu_B\right)$, $\omega$ is in units of $\Gamma/\hbar$ and spectral functions are in units of $\hbar/\Gamma$.
}
\label{fig:spec_eq}
\end{figure}
\end{center}

We start by plotting the impurity  spectral function in equilibrium ($\phi=0$) for different magnetic fields $B$, see \fig{fig:spec_eq}. 
Most of our results are obtained for an  interaction of $U = 6$, corresponding to a  Kondo temperature of $T_K \approx 0.2$.  
The temperature is fixed to $T/T_K \approx 0.25$. 
At finite magnetic field, the spin degeneracy is lifted, resulting in different spectral functions for spin-up and spin-down electrons.  
At particle-hole symmetry they are related to each other, according to $A_\uparrow(\omega) = A_\downarrow(-\omega)$. 
Upon increasing the magnetic field, the Kondo resonance is suppressed and it broadens, similarly to the effect of a bias voltage, cf. \tcites{wi.me.94,le.sc.01,ro.kr.01,ko.sc.96,fu.ue.03,sh.ro.06,fr.ke.10,nu.he.12,ande.08,ha.he.07,co.gu.14,do.nu.14,do.ga.15,do.ga.16}. 
Furthermore, a magnetic field causes a shift $\omega_B$ of the Kondo resonance to higher energies in the spin-resolved spectral function $A_\downarrow$ and produces a splitting $\delta_A$ in the total spectral function  $\atot = \tfrac{1}{2}\left( A_\uparrow + A_\downarrow \right)$. 
This splitting starts at $B \gtrsim T_K$, see also \tcites{cost.00,he.ba.05,wr.ga.11}, and persists until the peaks merge with the Hubbard bands.
The position of the Kondo resonance in $A_\downarrow$ becomes $\approx B$ for large $B$, while for decreasing $B$ the ratio $\omega_B/B$ decreases (see \fig{fig:delta}), consistent with previous results, mainly on the Kondo model.\cite{me.wi.93,mo.we.00,cost.00,lo.di.01,cost.03,ba.us.10} 
Note that for large magnetic fields one has $\delta_A = 2\omega_B$, while for small magnetic fields $\delta_A$ is smaller, due to the overlap of the contributions from the two spin directions.

\begin{center}
\begin{figure}[htp]
\subfigure[]{
  \label{fig:spec_neq_spin}
  \begin{minipage}[b]{\pw\columnwidth}
    \centering \includegraphics[width=1\textwidth]{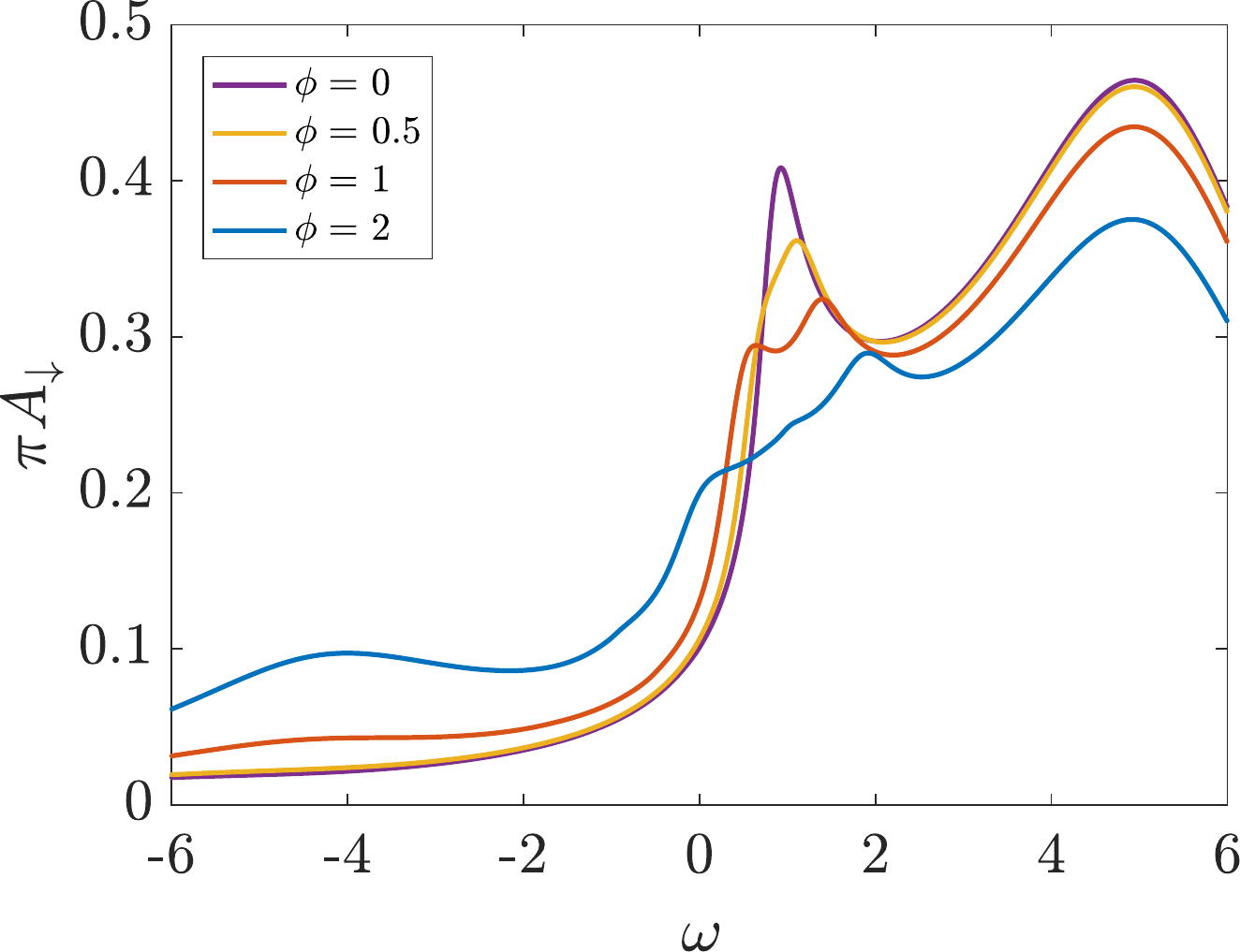}
  \end{minipage}
}\\
\subfigure[]{
  \label{fig:spec_neq_symm}
  \begin{minipage}[b]{\pw\columnwidth}
    \centering \includegraphics[width=1\textwidth]{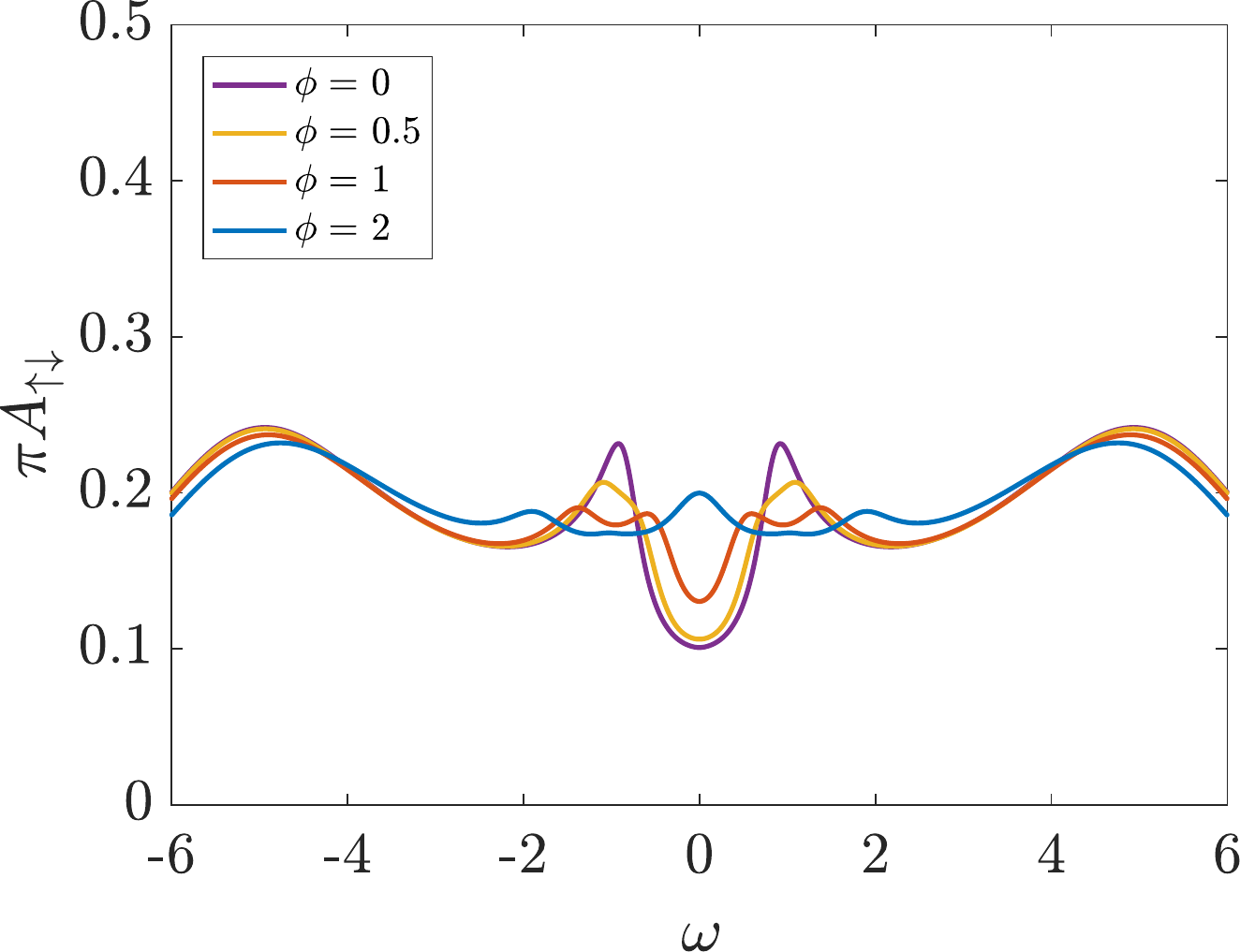}
  \end{minipage}
}\\
\caption{
Nonequilibrium spin-resolved \subref{fig:spec_neq_spin} and total \subref{fig:spec_neq_symm} impurity spectral function for different values of the bias voltage $\phi$ and fixed magnetic field $B=\Gamma/(|g|\mu_B)$; $T=0.05\Gamma/k_B \approx T_K/2$.
Note that $\phi$ is in units of $\Gamma/e$. 
Here a larger value of $U=8\Gamma$ is chosen, in order to resolve the four-peak structure in $\atot$.
}
\label{fig:spec_neq}
\end{figure}
\end{center} 

\begin{center}
\begin{figure}[htp]
\subfigure[]{
  \label{fig:curr_alone}
  \begin{minipage}[b]{\pw\columnwidth}
    \centering \includegraphics[width=1\textwidth]{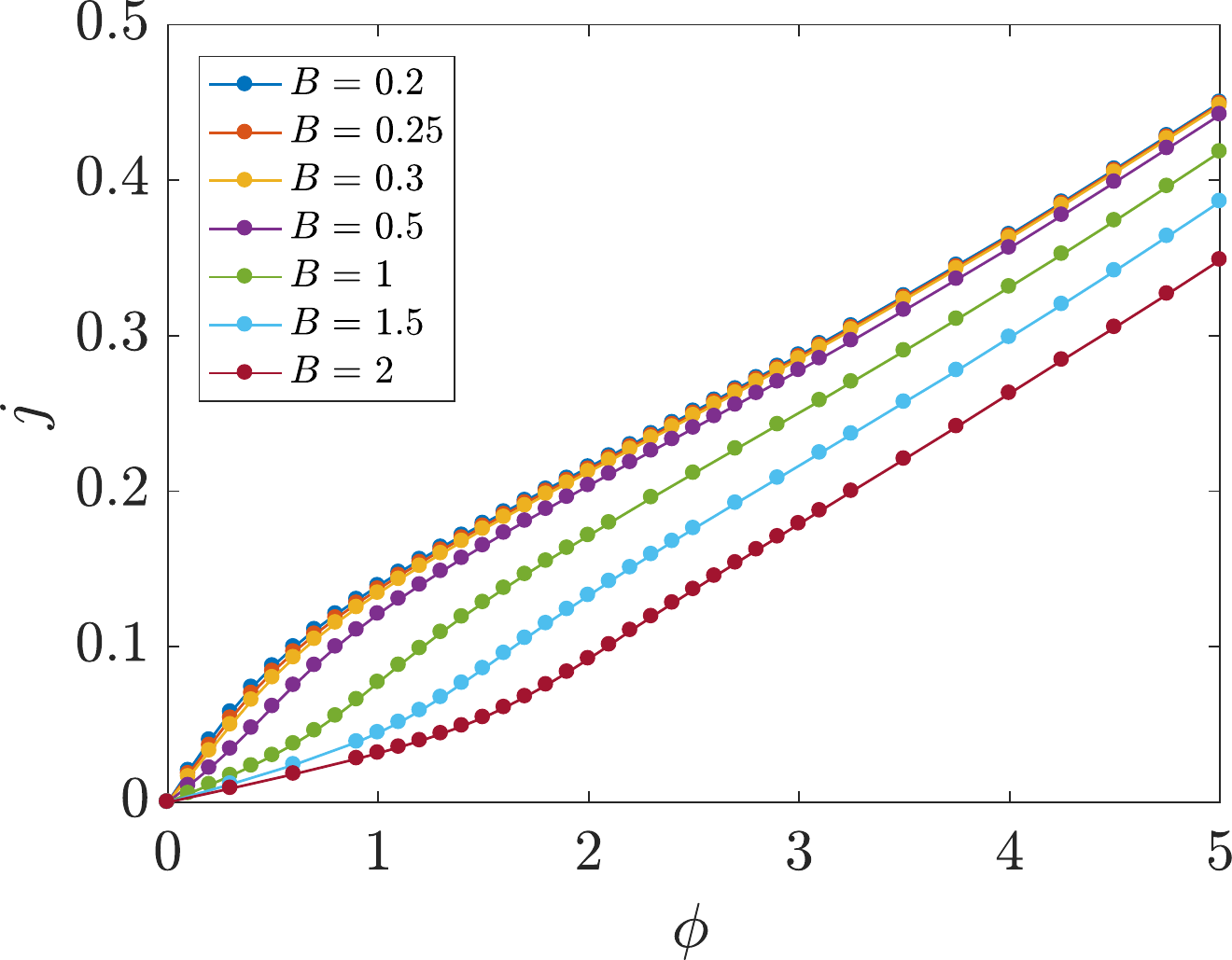}
  \end{minipage}
}\\
\subfigure[]{
  \label{fig:diffcond}
  \begin{minipage}[b]{\pw\columnwidth}
    \centering \includegraphics[width=1\textwidth]{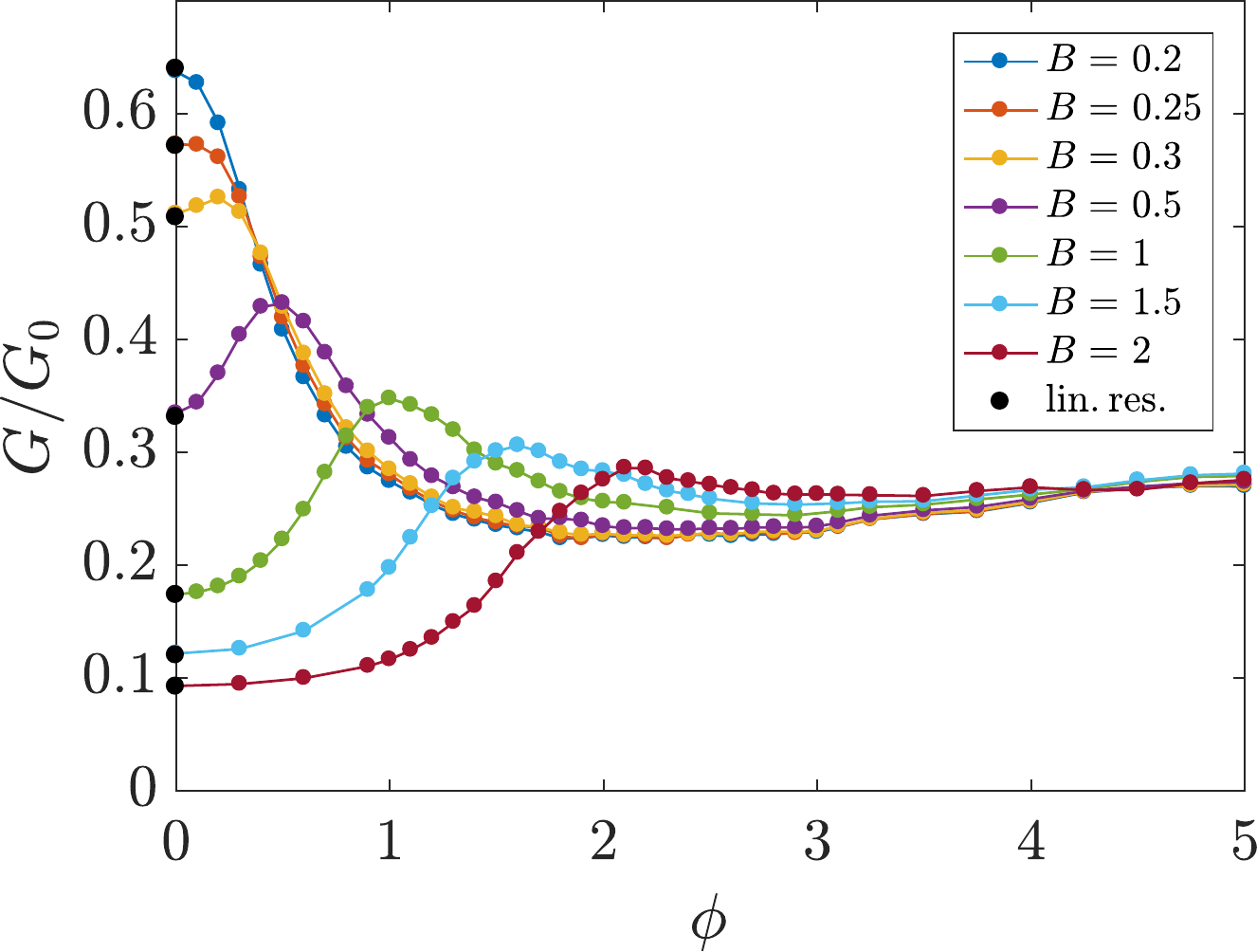}
  \end{minipage}
}\\
\caption{
\subref{fig:curr_alone} Current-voltage characteristic $j(\phi)$ and \subref{fig:diffcond} differential conductance $G(\phi)$ for different values of the magnetic field $B$.
$G$ is in units of $G_0 = G(T=0,\phi=0) = e^2/(\pi\hbar)$.
Parameters are as in \fig{fig:spec_eq}.
}
\label{fig:curr}
\end{figure}
\end{center} 

A similar splitting is produced by a bias voltage in the absence of a magnetic field,\cite{ande.08,do.nu.14,do.ga.15} so that it is interesting to study the combined behavior of the two effects.  
In the presence of both, a finite bias voltage and magnetic field, one would expect $4$ peaks in the total spectral function at $ \pm B\pm \phi/2$.
This has been observed within an equation of motion approach in \tcite{ba.us.10} (see also \tcite{ro.pa.03}). 
It is not easy to observe such a four-peak structure within a numerically controlled, nonperturbative approach. 
In our case, for $U=6$, the higher energy peaks merge with the Hubbard bands before the peaks are sufficiently far apart, so that they look more like shoulders than peaks.
For this reason, we investigate this effect for $U=8$.
\fig{fig:spec_neq_spin} shows the spin-resolved spectral functions $A_\downarrow(\omega)$ at $B=1$ for different bias voltages $\phi$ and $U=8$. 
At $\phi=0$ the position of the Kondo resonance $\omega_B$ is closer to $B$ than for the $U=6$ case, due to the fact that $T_K$ is smaller here.
As a result of the applied bias voltage the shifted Kondo resonance first acquires a broadening and then, starting from $\phi \approx 1$, it gets split. 
The two peaks have a distance of $\approx \phi$ as expected, but the splitting is not symmetric. 
The corresponding four-peak structure in the total spectral function can be seen in \fig{fig:spec_neq_symm} with split peaks at $\omega \simeq \pm B \pm \tfrac{\phi}{2}$, c.f. \onlinecite{ba.us.10}.

A more direct quantity to be measured experimentally is the differential conductance $G$ across the impurity. 
In \fig{fig:curr} we plot the current $j$ \subref{fig:curr_alone} as well as  $G$ \subref{fig:diffcond} as a function of the bias voltage for different values of $B$. 
The parameters are the same as in \fig{fig:spec_eq}.
To test the approaches, in  \fig{fig:diffcond_comp} we compare results from AMEA with the ones from the hybrid NRG-tDMRG calculation discussed in \se \ref{sec:frauke}. 
Results are essentially on top of each other.
The magnetic field affects the zero-bias peak in the conductance by first broadening it up to $B\gtrsim T_K$ and then producing a split,\cite{me.wi.93,ro.pa.03,he.ba.05,ande.08,sm.gr.13,ba.us.10} as observed experimentally.\cite{go.sh.98,cr.oo.98,kr.sh.11,kr.sh.12}
Notice that $\delta_G$, the splitting in $G$, starts at $B\approx 0.3$ and is slightly delayed in comparison to $\delta_A$, the splitting in $\atot(\omega)$, \fig{fig:spec_eq_symm}, which sets in at $B\approx 0.2$.
The reason for the delay in the splitting is the averaging of the spectral function in the current integral \eqref{eq:curr}, which smears out the effect of the split peaks. 
Since $G=G_\uparrow=G_\downarrow$ at particle-hole symmetry, $\phi_B$, the shift in the spin-resolved conductance $G_\downarrow$, exactly fulfills $\phi_B = \frac{\delta_G}{2}$, in contrast to its spectral counterpart, $\omega_B \geq \frac{\delta_A}{2}$.
On the other hand, the magnitude of the shift in $G$,
while becoming $\sim B$ for $B\gg T_K$, as shown in \fig{fig:delta}, it reaches this limit faster than the shift in $A_\downarrow(\omega)$.
In fact, \fig{fig:delta} suggests that, within the error bars~\footnote{
The error bars are rough estimates of the error in the numerical derivative of $j$ used to determine $G$. 
They are calculated under the conservative assumption that the turning points in $j(\phi)$, which determine the maxima in $G$, lie at most one voltage point off the calculated value.
}
$\phi_B$ becomes $\sim B$ as soon as it shows up, in contrast to $\omega_B$.
This is consistent with experiments,\cite{ra.bu.94,kr.sh.11,cr.oo.98} which indicate a strictly linear behavior.
At $\phi \gg B$ but smaller than the bandwidth the differential conductance reaches a $B$-independent value of $G \approx 0.27\,G_0$.

\begin{figure}[h]
\includegraphics[width=\pw\columnwidth]{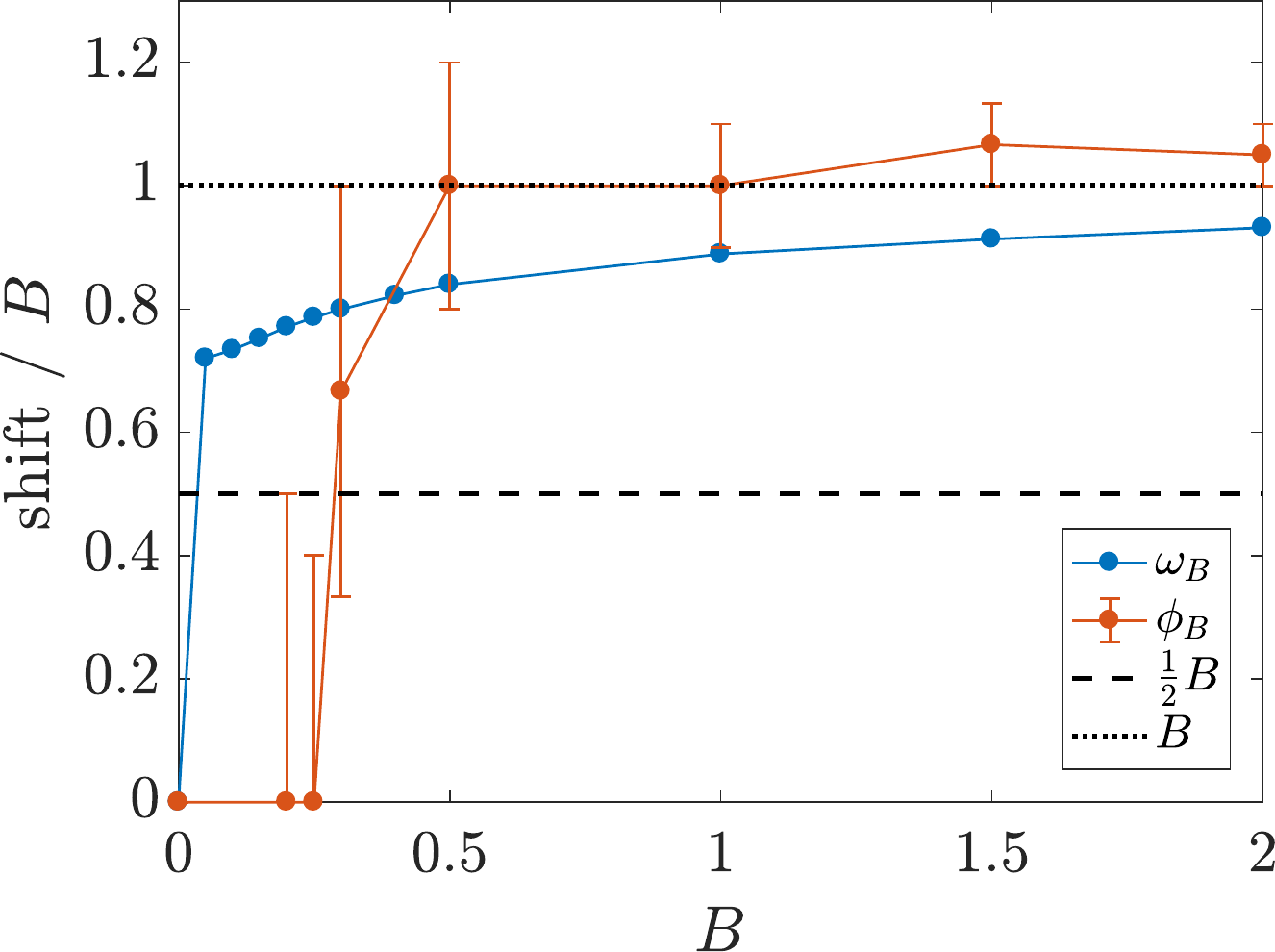}
\caption{
Shift $\phi_B$ of the conductance peak (in \fig{fig:diffcond}) and $\omega_B$ of the equilibrium spectral function (in \fig{fig:spec_eq_spin}) divided by the magnetic field $B$ plotted as a function of $B$.
Parameters are as in \fig{fig:spec_eq}.
}
\label{fig:delta}
\end{figure} 

\begin{center}
\begin{figure}[h]
\subfigure[]{
  \label{fig:mag_alone}
  \begin{minipage}[b]{\pw\columnwidth}
    \centering \includegraphics[width=1\textwidth]{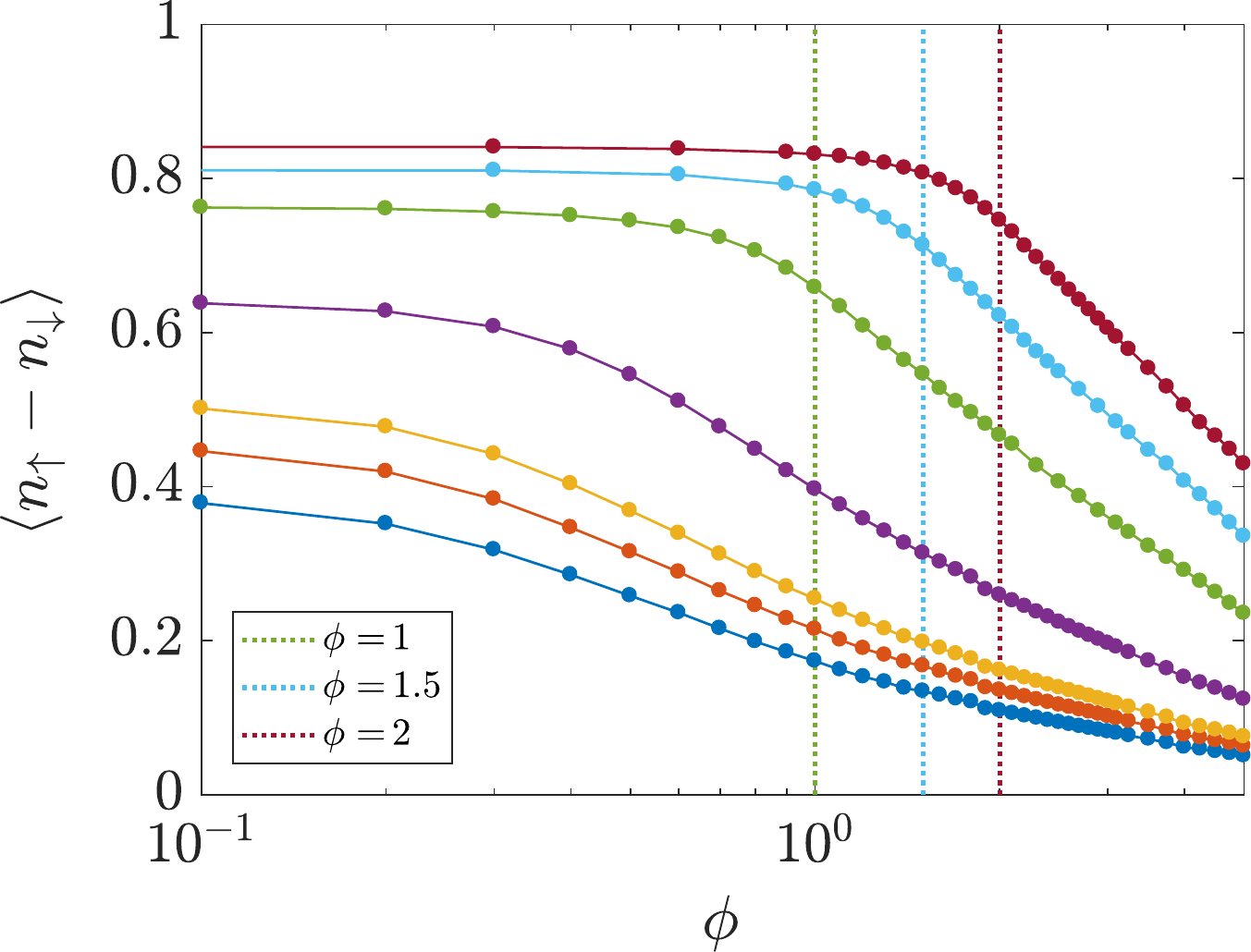}
  \end{minipage}
}\\
\subfigure[]{
  \label{fig:docc}
  \begin{minipage}[b]{\pw\columnwidth}
    \centering \includegraphics[width=1\textwidth]{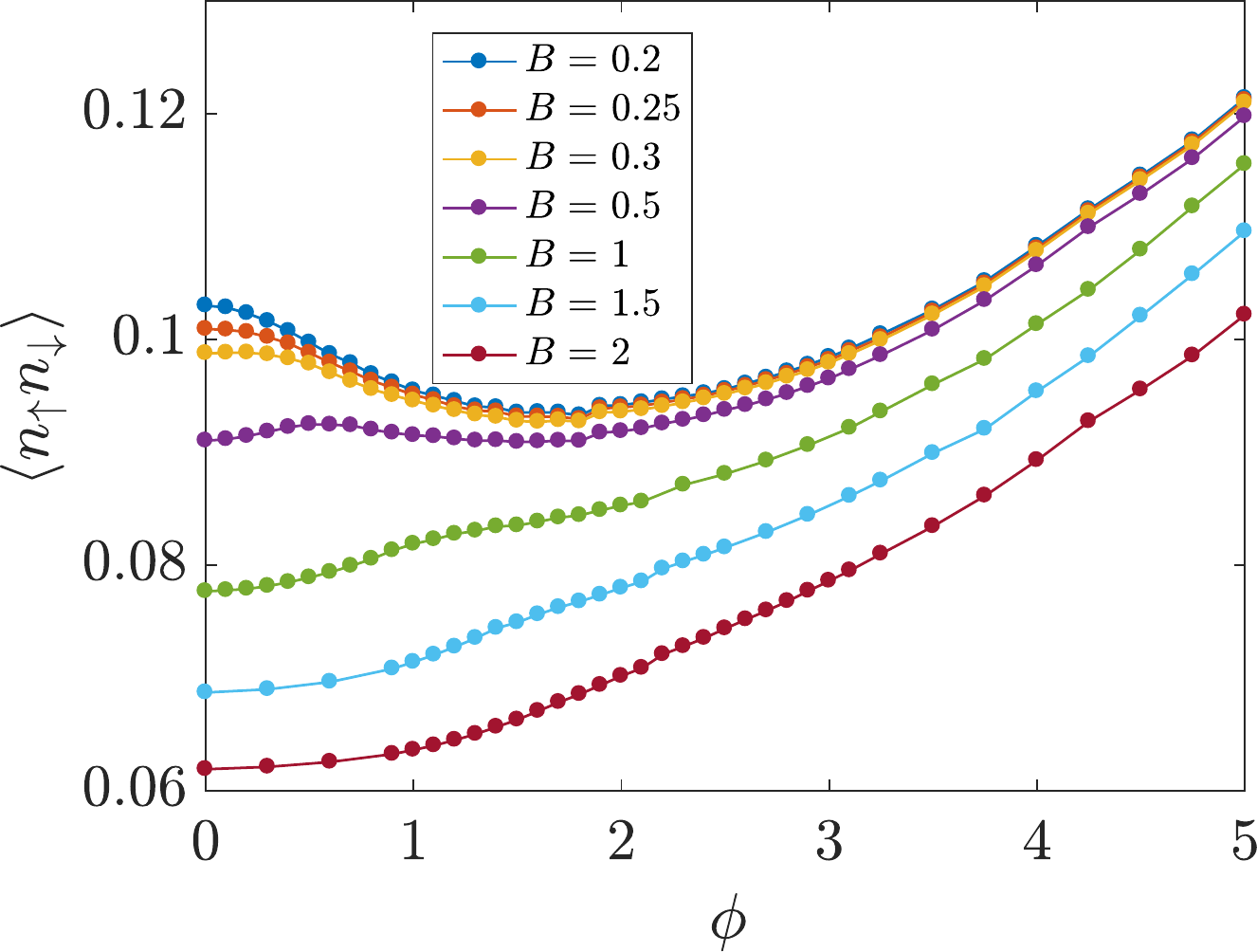}
  \end{minipage}
}\\
\caption{
\subref{fig:mag_alone} Magnetization and \subref{fig:docc} double occupancy as a function of the bias voltage $\phi$ for different values of the magnetic field $B$.
\subref{fig:docc} shares its legend with \subref{fig:mag_alone}. Dotted lines in \subref{fig:mag_alone} correspond to $\phi = B$.
Parameters are as in \fig{fig:spec_eq}.
}
\label{fig:mag}
\end{figure}
\end{center} 

\fig{fig:mag_alone} shows the magnetization $\braket{n_\uparrow - n_\downarrow}$ and \ref{fig:docc} the double occupancy $\braket{n_\uparrow n_\downarrow}$ at the impurity in dependence of the bias voltage for different magnetic fields.
At large magnetic fields $B \gg T_K$ the magnetization shows a plateau for $\phi \lesssim B$ followed by a logarithmic decrease (straight lines in \fig{fig:mag_alone}), in agreement with previous results, cf. \tcite{ro.pa.03}. 
At small magnetic fields $B \lesssim T_K$ it starts to decrease for $\phi \approx T_K$. 
Again, we find a very good agreement between AMEA and NRG-tDMRG, see \fig{fig:mag_comp}. 
For small magnetic fields the double occupancy has a minimum at $\phi \approx 2$, which seems to be independent of $T_K$, cf. \tcite{di.sc.13}. 
This minimum vanishes at larger magnetic fields as the Zeeman splitting of the local level increases and hence presumably is governed by charge fluctuations.

\begin{center}
\begin{figure}[htp]
\subfigure[]{
  \label{fig:spec_eq_comp}
  \begin{minipage}[b]{\pw\columnwidth}
    \centering \includegraphics[width=1\textwidth]{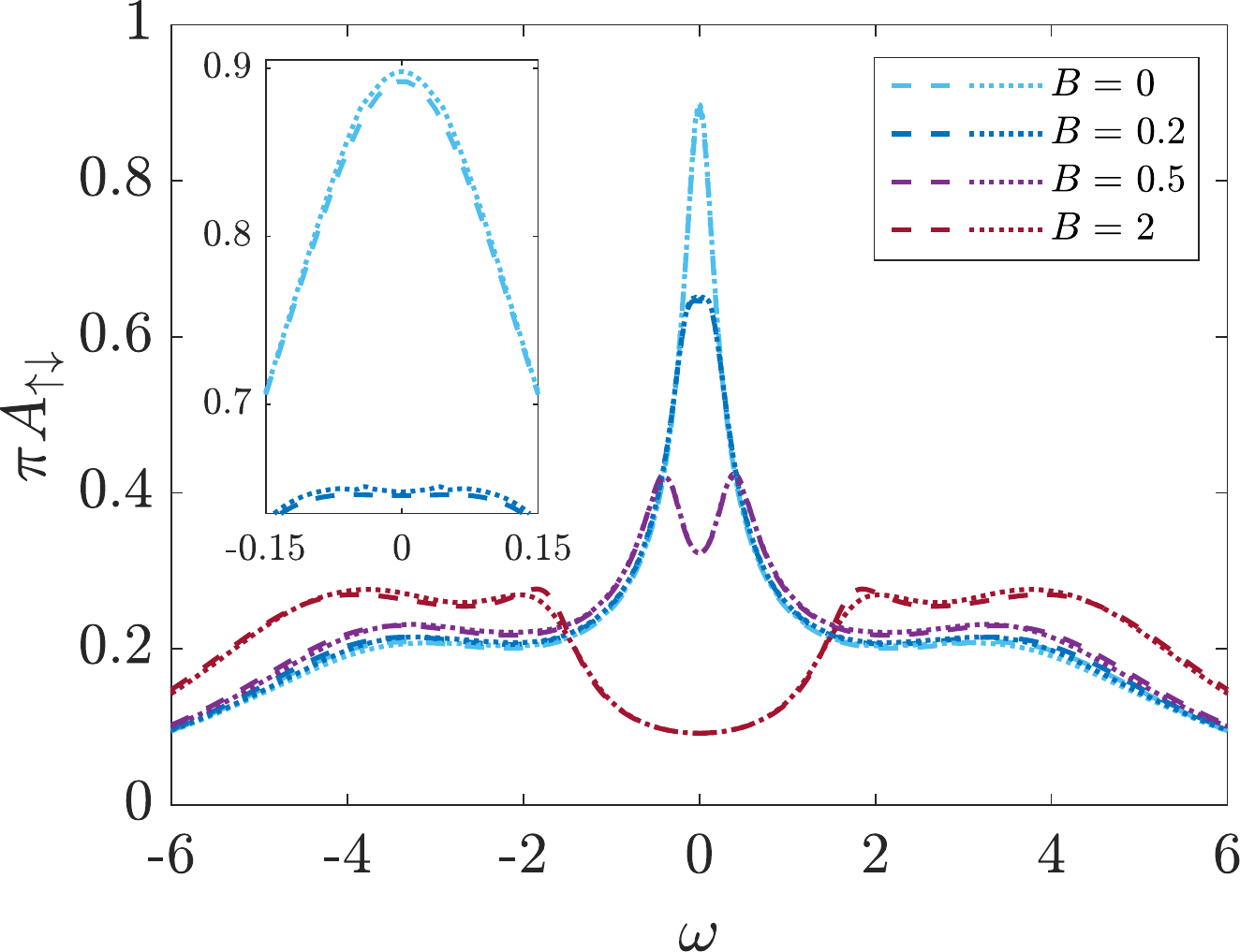}
  \end{minipage}
}\\
\subfigure[]{
  \label{fig:diffcond_comp}
  \begin{minipage}[b]{\pw\columnwidth}
    \centering \includegraphics[width=1\textwidth]{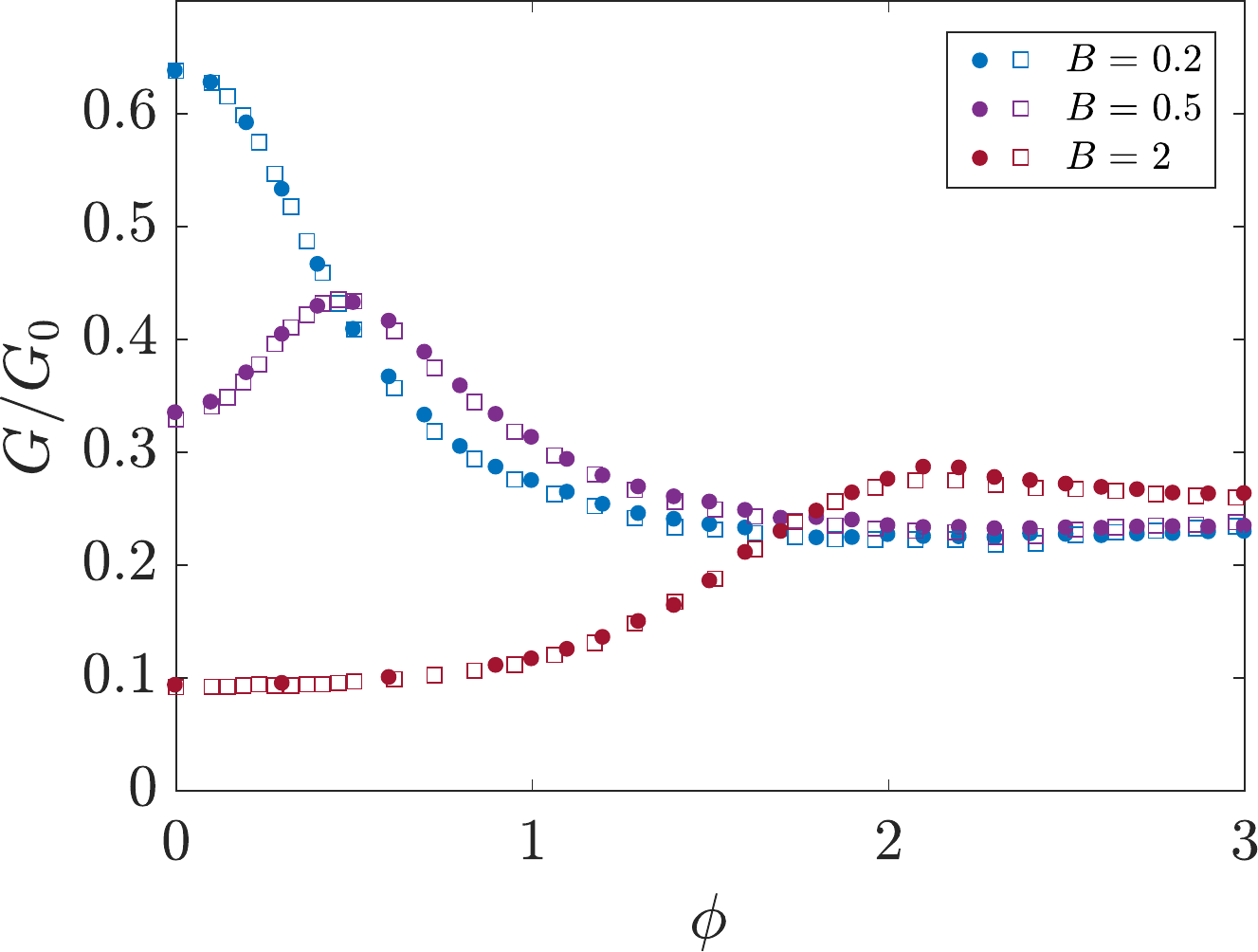}
  \end{minipage}
}\\
\subfigure[]{
  \label{fig:mag_comp}
  \begin{minipage}[b]{\pw\columnwidth}
    \centering \includegraphics[width=1\textwidth]{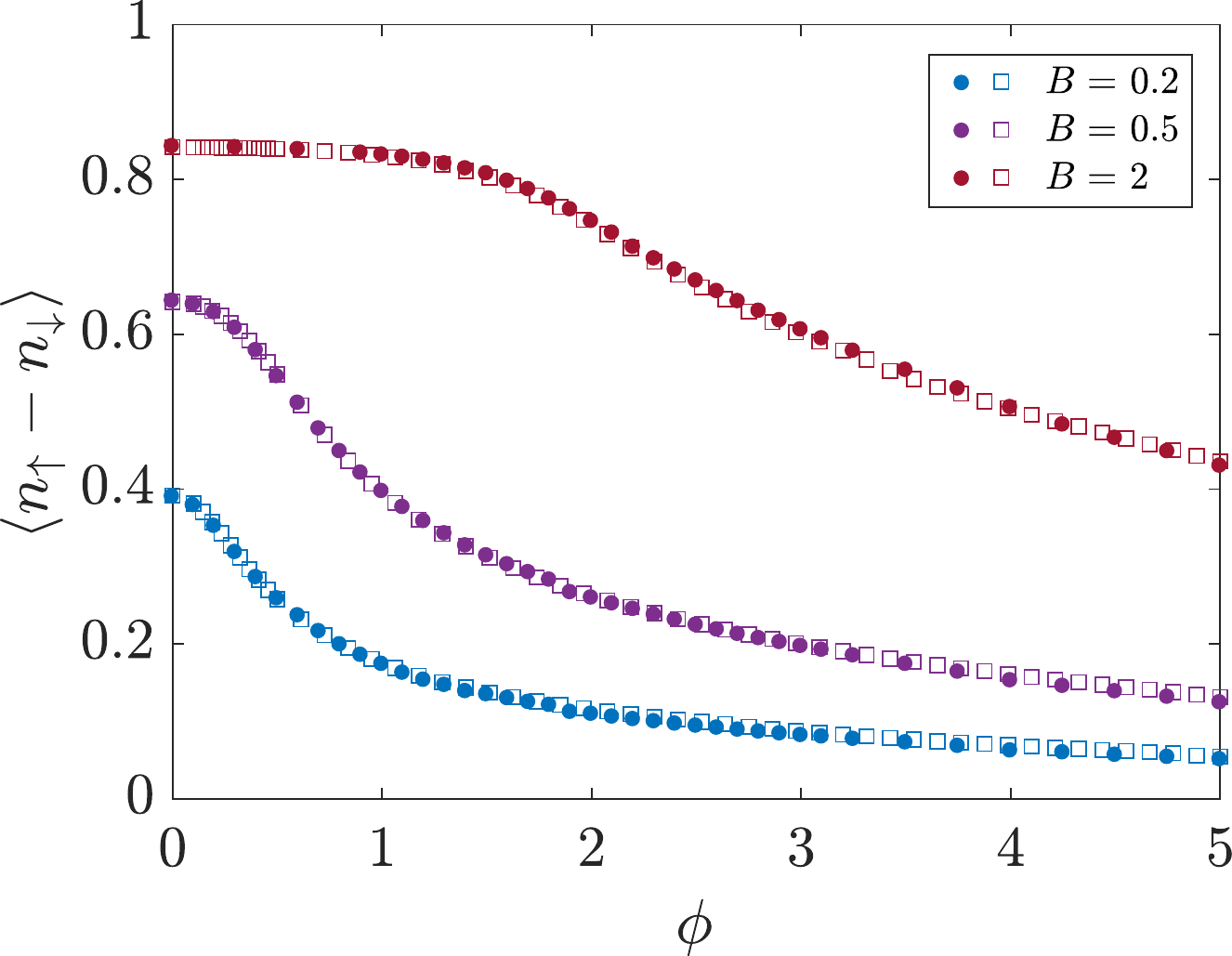}
  \end{minipage}
}\\
\caption{
Comparison of AMEA with NRG~\cite{we.de.07} and NRG-tDMRG~\cite{sc.we.17u}.
\subref{fig:spec_eq_comp} Equilibrium total impurity spectral function $\atot(\omega)$, \subref{fig:diffcond_comp} differential conductance $G(\phi)$ and \subref{fig:mag_comp} magnetization $\braket{n_\uparrow - n_\downarrow}(\phi)$ for different values of $B$.  
Dashed lines and circles correspond to AMEA, dotted lines to NRG and squares to NRG-tDMRG.
Parameters are as in \fig{fig:spec_eq}.
}
\label{fig:comp}
\end{figure}
\end{center} 

In \fig{fig:comp} we display a comparison of results obtained within AMEA (dashed lines and circles) with results from NRG ((a) dotted lines) and the hybrid NRG-tDMRG scheme discussed in \se \ref{sec:frauke} ((b,c) squares).
Equilibrium spectral functions \subref{fig:spec_eq_comp}, differential conductance \subref{fig:diffcond_comp} and magnetization \subref{fig:mag_comp} curves at different magnetic fields agree remarkably well between the two approaches.
One can only see small deviations in the spectral functions at high energies, due to the logarithmic discretization in NRG, which makes it less accurate in this energy region.
The inset in \subref{fig:spec_eq_comp} shows a zoom around $\omega=0$, where NRG is known to produce essentially exact  results. 
In this region the two spectral functions deviate by less than $1\%$.
The differential conductance at finite bias, being evaluated  from finite current differences (see \se \ref{sec:Keldysh}) in both approaches, is, in principle, more prone to errors.
Nontheless, the results lie essentially on top of each other.
On the other hand, as remarked in \se \ref{sec:frauke}, the magnetization from the NRG-tDMRG scheme is not fully converged to the steady state and the data have been extrapolated assuming an exponential decay of the occupancy $\braket{n_\sigma}(t)$.
For this reason, at high voltages, we can see that the values for the magnetization lie slightly above the AMEA results.
While it is, in principle, possible to calculate spectral functions within the NRG-tDMRG scheme, it is unclear at the moment, whether this is numerically feasible. 
For this reason, we don't provide a comparison between the two approaches in \fig{fig:comp}.

\section{Summary and Conclusions}
\label{sec:conclusion}

In this paper, we studied the Anderson impurity model out of equilibrium under the influence of a bias voltage $\phi$ and a magnetic field $B$.
In particular, we addressed the issue of the different behavior of the shift of the Kondo peak in the impurity spectral function and the one in the conductance anomaly as a function of the magnetic field. 
We also presented explicitly results for the spectral function showing a four-peak structure resulting from the combined effects of $B$ and $\phi$.

Our results agree with previous theoretical and experimental results in the known limits $B \ll T_K$ and $B \gg T_K$, while our approach allows us to access the intermediate regime $B, \phi \gtrsim T_K$ as well. 
The key aspect of our  auxiliary master equation approach~\cite{ar.kn.13,do.nu.14,do.ga.15,do.so.17} is that we can obtain very accurate results also for the spectral functions out of equilibrium, which is difficult by other methods. 
The accuracy of our results  in the parameter regime we considered is confirmed by an excellent comparison of spectral functions with  NRG at $\phi=0$ (up to frequencies for which NRG is supposed to yield correct results), and of expectation values with  a recently introduced hybrid NRG-tDMRG scheme~\cite{sc.we.17u} at finite bias voltages.

The two approaches adopted here, AMEA and the NRG-tDMRG scheme, deal with the challenge of describing the long time behavior of the nonequilibrium SIAM in a different manner. 
While AMEA explicitly describes an open quantum system and thus is not restricted to finite time scales, the quench approach renormalizes the problem down to the relevant energy scale.
In addition, AMEA is able to evaluate the impurity spectral function. 
While, in principle, this is also possible in the NRG-tDMRG approach, from a numerical point of view, it would be more costly.
Therefore, it is unclear at the moment, whether it is realizable in practice. 
Also for the magnetization AMEA was able to achieve better convergence, especially at high voltages.

In summary, it is convenient to use AMEA, whenever very long time scales are needed, or when information over the full energy range is required, as it is the case in the determination of spectral functions. 
For example, AMEA is an interesting tool for DMFT in nonequilibrium, where spectral functions are needed explicitly.\cite{ar.kn.13,ti.do.15,ti.do.16,do.ti.16,ti.do.17,so.do.17u}
On the other hand, the NRG-tDMRG approach is more flexible with respect to the parameter regime, as it uses an explicit renormalization of the impurity. 
In particular, it has proven to be able to describe very strong interactions such as $U/\Gamma=12$ and zero temperature $T=0$, see \tcite{sc.we.17u}.  
AMEA can deal with interactions of the same strength and temperatures down to $T \sim T_K/10$.\cite{do.ga.15}
Much larger values of $U$ and/or much lower in $T$ are not reachable at the moment, since we are limited in the number of bath sites.~\footnote{
Work is in progress on improving the Lindblad solver and achieve larger $N_B$. 
Notice that the accuracy increases exponentially with $N_B$.
}
This is also the reason, why we could not accurately check the well-known $\sim [\ln{\left(B/T_K\right)}]^{-2}$ behavior of $A_{\downarrow}(\omega=0)$ for $B\gg T_K$, \tcite{ro.co.03} in equilibrium.
Our results may be consistent with a logarithmic asymptotics, but, in order to reliably confirm this behavior, we need to consider magnetic fields that are orders of magnitude larger and at the same time $\ll U$. 
Therefore, at the moment, it may be preferable to use the NRG-tDMRG quench approach, whenever it gets crucial to work in the scaling limit and for very low values of the bias voltage.

The only approximation in AMEA consists in replacing the physical bath hybridization function $\underline{\Delta}$ with an auxiliary one $\underline{\Delta}_{aux}$, so that the accuracy depends on the difference between the two functions.
Of course, the corresponding error in the calculated results, e.g. the spectral function, is expected to be strongly frequency dependent, so that regions around the Fermi energies are probably more strongly affected. 
More specifically, due to the fact that at zero bias the Kondo scale depends exponentially on the $\omega=0$ DOS, one may expect a corresponding exponential error in this scale. 
This is probably not yet the case at these moderate values of $U/\Gamma \lesssim 8$ used here, as can be deduced from our results in \tcite{do.ga.15}. 
For larger $U$ (and more bath sites), the way to avoid this exponential problem could be to carry out the fit by constraining $\Im \Delta^R_{aux}$ to coincide with $\Im \Delta^R$ at $\omega = \mu_{R/L}$, or in any case require that the fit becomes more accurate around these points.

\begin{acknowledgments}

We would like to thank 
M. Sorantin, 
I. Titvinidze
and
W. von der Linden 
for fruitful discussions.
D.F., A.D. and E.A. were supported by the Austrian Science Fund (FWF) within the projects P26508 and F41 (SFB ViCoM), as well as NaWi Graz.
F.S. and J.v.D. were supported by the German-Israeli-Foundation through I-1259-303.10 and NIM. 
The calculations were partly performed on the D-Cluster Graz and on the VSC-3 cluster Vienna.

\end{acknowledgments}

\newpage

\begin{thebibliography}{10}

\bibitem{ra.bu.94}
D.~C. Ralph and R.~A. Buhrman, Phys. Rev. Lett. {\bf 72},  3401  (1994).

\bibitem{kr.sh.11}
A.~V. Kretinin, H. Shtrikman, D. Goldhaber-Gordon, M. Hanl, A. Weichselbaum, J.
  von Delft, T. Costi, and D. Mahalu, Phys. Rev. B {\bf 84},  245316  (2011).

\bibitem{kr.sh.12}
A.~V. Kretinin, H. Shtrikman, and D. Mahalu, Phys. Rev. B {\bf 85},  201301
  (2012).

\bibitem{go.go.98}
D. Goldhaber-Gordon, J. G{\"o}res, M.~A. Kastner, H. Shtrikman, D. Mahalu, and
  U. Meirav, Phys. Rev. Lett. {\bf 81},  5225  (1998).

\bibitem{go.sh.98}
D. Goldhaber-Gordon, H. Shtrikman, D. Mahalu, D. Abusch-Magder, U. Meirav, and
  M.~A. Kastner, Nature (London) {\bf 391},  156  (1998).

\bibitem{fe.ar.17}
M. Ferrier, T. Arakawa, T. Hata, R. Fujiwara, R. Delagrange, R. Deblock, Y.
  Teratani, R. Sakano, A. Oguri, and K. Kobayashi, Phys. Rev. Lett. {\bf 118},
  196803  (2017).

\bibitem{zh.ka.13}
Y. hui Zhang, S. Kahle, T. Herden, C. Stroh, M. Mayor, U. Schlickum, M. Ternes,
  P. Wahl, and K. Kern, Nature Communications {\bf 4},  2110  (2013).

\bibitem{am.ge.05}
S. Amasha, I.~J. Gelfand, M.~A. Kastner, and A. Kogan, Phys. Rev. B {\bf 72},
  045308  (2005).

\bibitem{me.wi.93}
Y. Meir, N.~S. Wingreen, and P.~A. Lee, Phys. Rev. Lett. {\bf 70},  2601
  (1993).

\bibitem{ba.us.10}
C.~A. Balseiro, G. Usaj, and M.~J. Sanchez, Journal of Physics: Condensed
  Matter {\bf 22},  425602  (2010).

\bibitem{he.ba.05}
A.~C. Hewson, J. Bauer, and A. Oguri, Journal of Physics: Condensed Matter {\bf
  17},  5413  (2005).

\bibitem{ro.pa.03}
A. Rosch, J. Paaske, J. Kroha, and P. W\"olfle, Phys. Rev. Lett. {\bf 90},
  076804  (2003).

\bibitem{re.pl.14}
F. Reininghaus, M. Pletyukhov, and H. Schoeller, Phys. Rev. B {\bf 90},  085121
   (2014).

\bibitem{ande.08}
F.~B. Anders, Phys. Rev. Lett. {\bf 101},  066804  (2008).

\bibitem{sm.gr.13}
S. Smirnov and M. Grifoni, New Journal of Physics {\bf 15},  073047  (2013).

\bibitem{mo.we.00}
J.~E. Moore and X.-G. Wen, Phys. Rev. Lett. {\bf 85},  1722  (2000).

\bibitem{cost.00}
T.~A. Costi, Phys. Rev. Lett. {\bf 85},  1504  (2000).

\bibitem{cost.03}
T.~A. Costi,  in {\em Concepts in Electron Correlation}, Vol.~110 of {\em NATO
  science series II:}, edited by A.~C. Hewson and V. Zlatic (Springer Science
  {\&} Business Media, Dodrecht, 2003), p.\ 247.

\bibitem{ro.co.03}
A. Rosch, T.~A. Costi, J. Paaske, and P. W\"olfle, Phys. Rev. B {\bf 68},
  014430  (2003).

\bibitem{me.wi.92}
Y. Meir and N.~S. Wingreen, Phys. Rev. Lett. {\bf 68},  2512  (1992).

\bibitem{do.ga.15}
A. Dorda, M. Ganahl, H.~G. Evertz, W. von~der Linden, and E. Arrigoni, Phys.
  Rev. B {\bf 92},  125145  (2015).

\bibitem{sc.we.17u}
F. Schwarz, I. Weymann, J. von Delft and A. Weichselbaum, arXiv:1708.06315 (unpublished).

\bibitem{ha.ja}
H. Haug and A.-P. Jauho, {\em Quantum Kinetics in Transport and Optics of
  Semiconductors} (Springer, Heidelberg, 1998).

\bibitem{do.nu.14}
A. Dorda, M. Nuss, W. von~der Linden, and E. Arrigoni, Phys. Rev. B {\bf 89},
  165105  (2014).

\bibitem{do.so.17}
A. Dorda, M. Sorantin, W. von~der Linden, and E. Arrigoni, New J. Phys. {\bf
  19},  063005  (2017).

\bibitem{ar.kn.13}
E. Arrigoni, M. Knap, and W. von~der Linden, Phys. Rev. Lett. {\bf 110},
  086403  (2013).

\bibitem{pros.08}
T. Prosen, New J. Phys. {\bf 10},  043026  (2008).

\bibitem{pros.10}
T. Prosen, Journal of Statistical Mechanics: Theory and Experiment {\bf 2010},
  P07020  (2010).

\bibitem{bo.sa.08}
E. Boulat, H. Saleur, and P. Schmitteckert, Phys. Rev. Lett. {\bf 101},  140601
   (2008).

\bibitem{he.fe.09}
F. Heidrich-Meisner, A.~E. Feiguin, and E. Dagotto, Phys. Rev. B {\bf 79},
  235336  (2009).

\bibitem{wi.ke.75}
K.~G. Wilson, Rev. Mod. Phys. {\bf 47},  773  (1975).

\bibitem{vida.04}
G. Vidal, Phys. Rev. Lett. {\bf 93},  040502  (2004).

\bibitem{da.ko.04}
A.~J. Daley, C. Kollath, U. Schollw{\"o}ck, and G. Vidal, J. Stat. Mech. {\bf
  2004},  P04005  (2004).

\bibitem{wh.fe.04}
S.~R. White and A.~E. Feiguin, Phys. Rev. Lett. {\bf 93},  076401  (2004).

\bibitem{scho.11}
U. Schollw{\"o}ck, Annals of Physics {\bf 326},  96   (2011).

\bibitem{ta.um.75}
Y. Takahashi and H. Umezawa, Collective Phenomena {\bf 2},  55–80  (1975).

\bibitem{das.00}
A. Das,  in {\em Quantum Field Theory - A 20th Century Profile}, edited by
  A.~N. Mitra (Hindustan Book Agency, New Delhi, 2000), pp.\ 383--411.

\bibitem{ve.in.15}
I. de~Vega and M.-C. Ba{\~n}uls, Phys. Rev. A {\bf 92},  052116  (2015).

\bibitem{wi.me.94}
N.~S. Wingreen and Y. Meir, Phys. Rev. B {\bf 49},  11040  (1994).

\bibitem{le.sc.01}
E. Lebanon and A. Schiller, Phys. Rev. B {\bf 65},  035308  (2001).

\bibitem{ro.kr.01}
A. Rosch, J. Kroha, and P. W{\"o}lfle, Phys. Rev. Lett. {\bf 87},  156802
  (2001).

\bibitem{ko.sc.96}
J. K{\"o}nig, J. Schmid, H. Schoeller, and G. Sch{\"o}n, Phys. Rev. B {\bf 54},
   16820  (1996).

\bibitem{fu.ue.03}
T. Fujii and K. Ueda, Phys. Rev. B {\bf 68},  155310  (2003).

\bibitem{sh.ro.06}
N. Shah and A. Rosch, Phys. Rev. B {\bf 73},  081309  (2006).

\bibitem{fr.ke.10}
P. Fritsch and S. Kehrein, Phys. Rev. B {\bf 81},  035113  (2010).

\bibitem{nu.he.12}
M. Nuss, C. Heil, M. Ganahl, M. Knap, H.~G. Evertz, E. Arrigoni, and W. von~der
  Linden, Phys. Rev. B {\bf 86},  245119  (2012).

\bibitem{ha.he.07}
J.~E. Han and R.~J. Heary, Phys. Rev. Lett. {\bf 99},  236808  (2007).

\bibitem{co.gu.14}
G. Cohen, E. Gull, D.~R. Reichman, and A.~J. Millis, Phys. Rev. Lett. {\bf
  112},  146802  (2014).

\bibitem{do.ga.16}
A. Dorda, M. Ganahl, S. Andergassen, W. von~der Linden, and E. Arrigoni, Phys.
  Rev. B {\bf 94},  245125  (2016).

\bibitem{wr.ga.11}
C.~J. Wright, M.~R. Galpin, and D.~E. Logan, Phys. Rev. B {\bf 84},  115308
  (2011).

\bibitem{lo.di.01}
D.~E. Logan and N.~L. Dickens, Journal of Physics: Condensed Matter {\bf 13},
  9713  (2001).

\bibitem{cr.oo.98}
S. Cronenwett, T. Oosterkamp, and L. Kouwenhoven, Science {\bf 281},  540
  (1998).

\bibitem{di.sc.13}
A. Dirks, S. Schmitt, J.~E. Han, F. Anders, P. Werner, and T. Pruschke, EPL
  (Europhysics Letters) {\bf 102},  37011  (2013).

\bibitem{we.de.07}
A. Weichselbaum and J. von Delft, Phys. Rev. Lett. {\bf 99},  076402  (2007).

\bibitem{ti.do.15}
I. Titvinidze, A. Dorda, W. von~der Linden, and E. Arrigoni, Phys. Rev. B {\bf
  92},  245125  (2015).

\bibitem{ti.do.16}
I. Titvinidze, A. Dorda, W. von~der Linden, and E. Arrigoni, Phys. Rev. B {\bf
  94},  245142  (2016).

\bibitem{do.ti.16}
A. Dorda, I. Titvinidze, and E. Arrigoni, Journal of Physics: Conference Series
  {\bf 696},  012003  (2016).

\bibitem{ti.do.17}
I. Titvinidze, A. Dorda, W. von~der Linden, and E. Arrigoni, Phys. Rev. B {\bf
  96},  115104  (2017).

\bibitem{so.do.17u}
M.~E. Sorantin, A. Dorda, K. Held, and E. Arrigoni, arXiv:1708.05011
  (unpublished).

\end{thebibliography}

\end{document}